\documentclass[aps,prl,twocolumn,showkeys,superscriptaddress]{revtex4-1}

\usepackage{placeins}
\usepackage[artemisia]{textgreek}
\usepackage{amssymb}
\usepackage{epsfig}
\usepackage{amsmath}
\usepackage{graphicx}
\usepackage{dcolumn}
\usepackage{latexsym}
\usepackage{lipsum}  
\usepackage{color}
\usepackage{epstopdf}
\usepackage{subfigure}
\usepackage{nicefrac}
\usepackage{blindtext}
\usepackage[ulem=normalem]{changes}
\usepackage{float}
\usepackage{soul}
\usepackage[T1]{fontenc}
\usepackage[hidelinks]{hyperref}

\hypersetup{
    colorlinks,
    linkcolor={red!50!black},
    citecolor={blue!50!black},
    urlcolor={blue!80!black}
}

\newcommand*{ \citen}[1]{%
  \begingroup
    \romannumeral-`\x 
    \setcitestyle{numbers}%
     \cite{#1}%
  \endgroup   
}

\begin{document}

\title{Quench switching of Mn$_2$As}

\author{Kamil Olejník}
\affiliation{Institute of Physics, Czech Academy of Sciences, 162 00 Prague, Czech Republic}

\author{Zdeněk Kašpar}
\affiliation{Institute of Physics, Czech Academy of Sciences, 162 00 Prague, Czech Republic}
\affiliation{Faculty of Mathematics and Physics, Charles University, 121 16 Prague, Czech Republic}

\author{Jan Zubáč}
\affiliation{Institute of Physics, Czech Academy of Sciences, 162 00 Prague, Czech Republic}
\affiliation{Faculty of Mathematics and Physics, Charles University, 121 16 Prague, Czech Republic}

\author{Sjoerd Telkamp}
\affiliation{Solid State Physics Laboratory, ETH Zurich, 8093 Zurich, Switzerland}

\author{Andrej Farkaš}
\affiliation{Institute of Physics, Czech Academy of Sciences, 162 00 Prague, Czech Republic}
\affiliation{Faculty of Mathematics and Physics, Charles University, 121 16 Prague, Czech Republic}

\author{Dominik Kriegner}
\affiliation{Institute of Physics, Czech Academy of Sciences, 162 00 Prague, Czech Republic}

\author{Karel Výborný}
\affiliation{Institute of Physics, Czech Academy of Sciences, 162 00 Prague, Czech Republic}

\author{Jakub Železný}
\affiliation{Institute of Physics, Czech Academy of Sciences, 162 00 Prague, Czech Republic}

\author{Zbyněk Šobáň}
\affiliation{Institute of Physics, Czech Academy of Sciences, 162 00 Prague, Czech Republic}

\author{Peng Zeng}
\affiliation{ScopeM, ETH Zurich, 8093 Zurich, Switzerland}

\author{Tomáš Jungwirth}
\affiliation{Institute of Physics, Czech Academy of Sciences, 162 00 Prague, Czech Republic}

\author{Vít Novák}
\affiliation{Institute of Physics, Czech Academy of Sciences, 162 00 Prague, Czech Republic}

\author{Filip Krizek}
\email{krizekfi@fzu.cz}
\affiliation{Institute of Physics, Czech Academy of Sciences, 162 00 Prague, Czech Republic}

\date{\today}

\begin{abstract}

We demonstrate that epitaxial thin film antiferromagnet Mn$_2$As exhibits the quench-switching effect, which was previously reported only in crystallographically similar antiferromagnetic CuMnAs thin films. Quench switching in Mn$_2$As shows stronger increase in resistivity, reaching hundreds of percent at 5K, and significantly longer retention time of the metastable high-resistive state before relaxation towards the low-resistive uniform magnetic state. Qualitatively, Mn$_2$As and CuMnAs show analogous parametric dependence of the magnitude and relaxation of the quench-switching signal. Quantitatively, relaxation dynamics in both materials show direct proportionality to the Néel temperature. This confirms that the quench switching has magnetic origin in both materials. The presented results suggest that the antiferromagnets crystalizing in the Cu$_2$Sb structure are well suited for exploring and exploiting the intriguing physics of highly non-uniform magnetic states associated with the quench switching.

\end{abstract}
\maketitle

\section{Introduction}

In the past decade, the unique properties of antiferromagnetism inspired new research directions in condensed matter, especially with focus on manipulation of their Néel vector \cite{jungwirth2016antiferromagnetic,baltz2018antiferromagnetic,xiong2022antiferromagnetic,chen2024emerging}. Its orientation can be detected electrically, e.g.,~via anisotropic magnetoresistance \cite{marti2014room,wadley2016electrical}, spin-Hall magnetoresistance~\cite{Han2016}, or second-harmonic magnetoresistance~\cite{godinho2018}. However,
because of the relativistic nature of these phenomena the corresponding signals are weak, typically below one percent. 

We have previously demonstrated a strong and reproducible non-relativistic mechanism of manipulation of magnetic order in CuMnAs --- the quench switching \cite{kavspar2021quenching,zubavc2021hysteretic,surynek2024picosecond}. This effect is induced by electrical or optical pulses of length ranging from ms to fs, which temporarily heat the material above its Néel temperature (475~K in CuMnAs \cite{wadley2015antiferromagnetic}) and allow it to rapidly cool down. As a result the antiferromagnet remains frozen (quenched) in a metastable state, associated with a strong fragmentation of the antiferromagnetic order by a formation of nano-scale magnetic textures [5,6,7,9], which only slowly relaxes back towards the uniform equilibrium state of the antiferromagnet.
The metastable state features a significant increase of resistivity, which can be measured in a straightforward two-point or four-point in-plane resistor geometry. The quench-switching effect in CuMnAs has several intriguing aspects: (i) the high-resistive metastable states are multi-level, controllable by magnitude, length, delay or order of switching pulses; (ii) switching times can match characteristic $\mu$s/ns scales [5] of microcontrollers/microprocessors, as well as ps-scales [5,7] of THz or optical technologies; (iii) the change in resistivity at room temperature can reach ten percent (100\% at low temperature) with very low energy cost of $\approx$ aJ/nm$^3$ required for switching~\cite{wadley2016electrical}; 
(iv) the (sub)ps attempt rates and activation barriers $E/k_B \approx 30\times 300$~K result in temperature-dependent relaxation times spanning many orders of magnitude (from seconds at 300 K to hours at 230 K \cite{kavspar2021quenching}); (v) combination of the above features conforms to the requirements of analog time-dependent logic-in-memory (spiking neuromorphic) circuits \cite{surynek2024picosecond}. 
However, up to now the observation of the above phenomenology
of the quench switching remained constrained to a single antiferromagnetic material --- the epitaxial thin film of tetragonal CuMnAs. 

In this paper we report on epitaxial thin film Mn$_2$As grown on GaAs substrate, which exhibits a qualitatively analogous phenomenology of the quench-switching effect as in CuMnAs. Quantitatively, the magnitude and retention time of the switching signal is larger, in proportion to the hundred degrees higher N\'eel temperature in Mn$_2$As.

\begin{figure*}[ht!]
\vspace{0.2cm}
\includegraphics[scale=0.25]{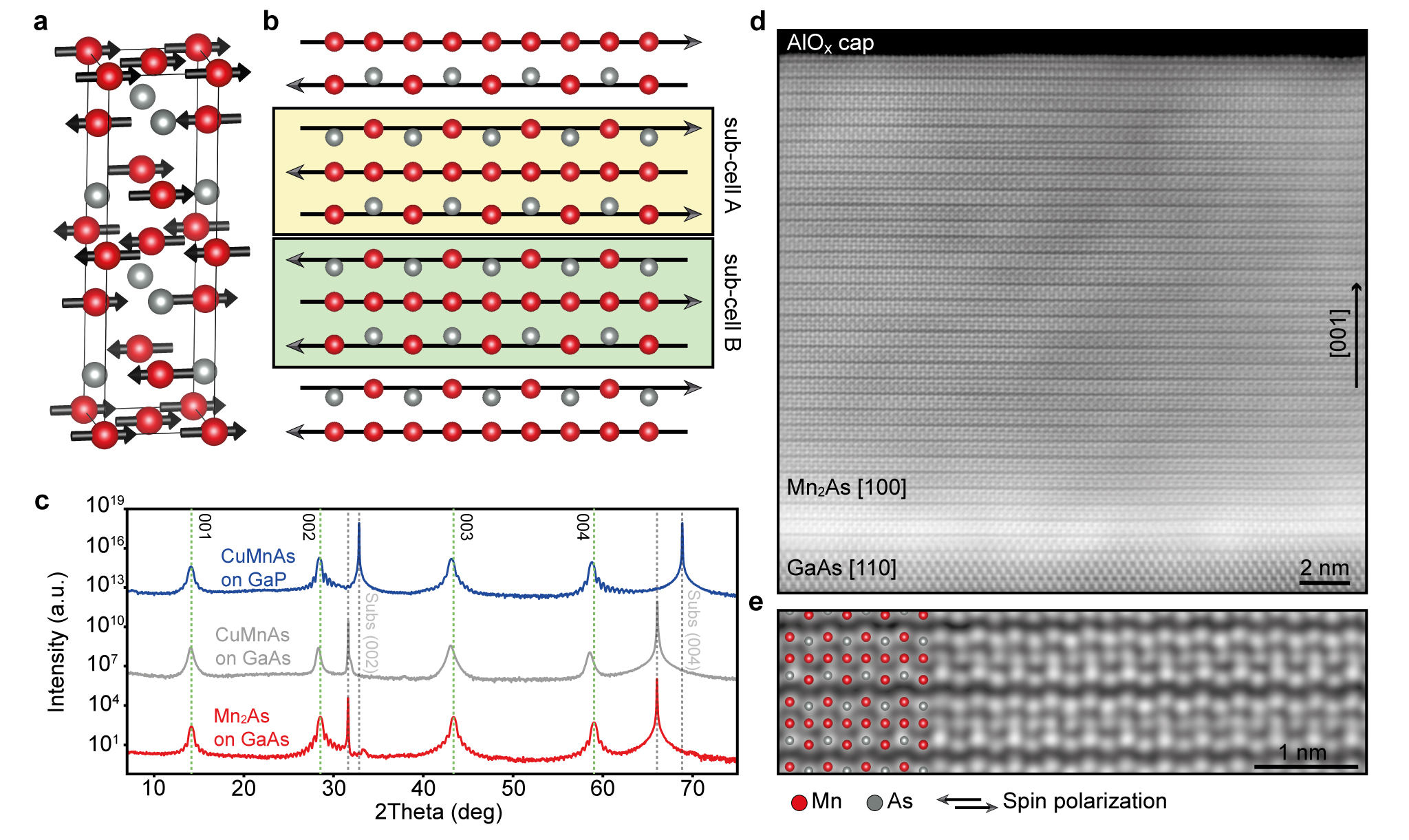}
\vspace{-0.4cm}
\caption{(a) 
Unit cell model of Mn$_{2}$As (Mn - red, As grey), where the arrows highlight the expected local moment orientation. (b) [110] projection model of a portion of Mn$_{2}$As where the antifferomagnetic compensation between magnetic sub-cells A and B is highlighted within the yellow and green rectangle. (c) XRD radial scan, demonstrating high quality of a 20 nm thick Mn$_{2}$As (red) film grown on GaAs together with comparison with CuMnAs grown on GaAs (grey) and GaP (blue) substrates. The green dashed lines highlight 
positions corresponding to the 00L Bragg peaks of Mn$_{2}$As and grey dashed lines the 002 and 004 Bragg peaks of the substrate. (d) HAADF-STEM projection along [100] direction of a 20 nm film of Mn$_{2}$As 
grown on a GaAs substrate. (e) HAADF-STEM zoom-in on a portion of the Mn$_{2}$As, with an overlay indicating the Mn (red) and As (grey) atoms.}
\label{fig1}
\end{figure*}

Mn$_2$As was selected from an in this context virtually unexplored material family of tetragonal 3d metal pnictides \cite{motizuki2010electronic}, which belong to the Cu$_2$Sb-like P4/\textit{nmm} space group and stabilize in a wide range of accessible magnetic orders \cite{zhang2013magnetic}. The previously synthesized bulk form of Mn$_2$As is known to be isomorphic to tetragonal CuMnAs, with Cu$_2$Sb-type P4/nmm crystal structure and lattice parameters $a=3.78\,$\AA $ $ and $c=6.28\,$\AA $ $ \cite{austin1962magnetic}, very close to $a=3.85\,$\AA$ $ and $c=6.28\,$\AA $ $ of CuMnAs. The antiferromagnetic magnetic structure of Mn$_{2}$As belongs to P$_{a}$nma magnetic space group, which has doubled unit-cell along the [100] direction, as illustrated in Fig.~\ref{fig1}a.

Beside the structural similarity between CuMnAs and Mn$_2$As and the fact that both are compensated collinear in-plane antiferromagnets, there are also important differences which 
are highly relevant for the comparison of the quench switching process in the two materials: (i) Mn$_2$As has higher Néel temperature of 570~K - 573~K \cite{yuzuri1960magnetic,austin1962magnetic}, approximately by 100 K higher than CuMnAs; (ii) both materials have a different magnetic symmetry: CuMnAs is \textit{PT} symmetric, while Mn$_2$As is \textit{P} symmetric with broken \textit{PT} symmetry; (iii) in Mn$_2$As, Mn/Cu anti-site defects do not exist, which in CuMnAs represent the most likely low-energy point defects \cite{maca2019tetragonal}.


\section{Material Growth and Properties}

The Mn$_{2}$As samples were grown on GaAs (001) substrates by molecular beam epitaxy (MBE). After desorption of the native oxide 
the GaAs wafer was cooled to 570$^{\circ}$C and a 100 nm thick GaAs buffer layer was grown. To suppress a possible current leakage to the GaAs substrate in electrical measurements, a 500 nm thick Al$_{0.4}$Ga$_{0.6}$As was grown as a second buffer layer in selected samples. After growth of the buffer, the sample was cooled down to 220$^{\circ}$C (measured by band edge spectroscopy \cite{novak2007substrate}) and 20, 50 or 78 nm thick Mn$_{2}$As films were grown at 0.024\AA/s. Finally, the Mn$_{2}$As film was capped with a 3 nm thick layer of Al (at substrate temperature below 10$^{\circ}$C) which, after oxidation at ambient conditions, forms a protective AlO$_\mathrm{x}$ layer. Maintaining the 2:1 ratio of the Mn and As fluxes turns out to be crucial for the successful growth, since only a slight deviation from the correct stoichiometry results in surface roughening and/or formation of ferromagnetic MnAs. In our case, the non-uniformity of Mn and/or As fluxes results in gradient of stoichoimetry accross the wafer, and Mn$_2$As with ideal stoichiometry grows only in a few-millimeter wide annular area of the 2" wafer (see SEM and RHEED images in the Supplementary information. Reference CuMnAs samples were grown under conditions given in Ref. \cite{krizek2020molecular}.

\begin{figure}[hbt!]
\vspace{0.2cm}
\includegraphics[scale=0.115]{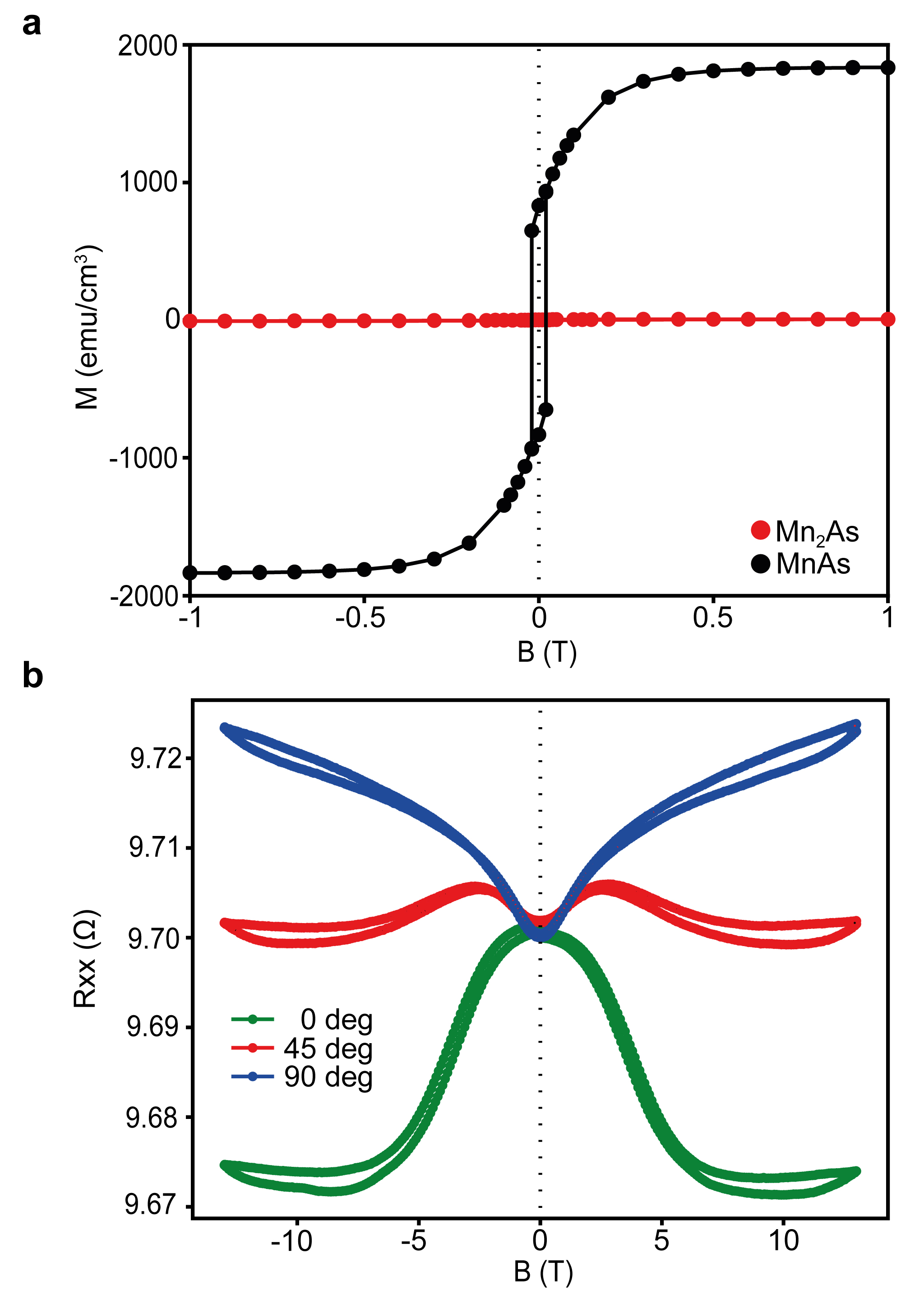}
\vspace{-0.4cm}
\caption{(a) Out-of-plane magnetization of Mn$_{2}$As and, for comparison, ferromagnetic MnAs thin films; the measurements were performed at 10~K. Data are shown after subtraction of the diamagnetic contribution of the substrate. (b) Dependency of longitudinal resistance of Mn$_{2}$As in a Hall-bar geometry on in-plane external magnetic field. 0~deg, 45~deg and 90~deg indicate angles between the magnetic field and the current direction; the Hall-bar was oriented along the [100] direction of Mn$_2$As, the measurements were performed at 80~K.}
\label{fig2}
\end{figure}

\begin{figure*}[hbt!]
\vspace{0.2cm}
\includegraphics[scale=0.25]{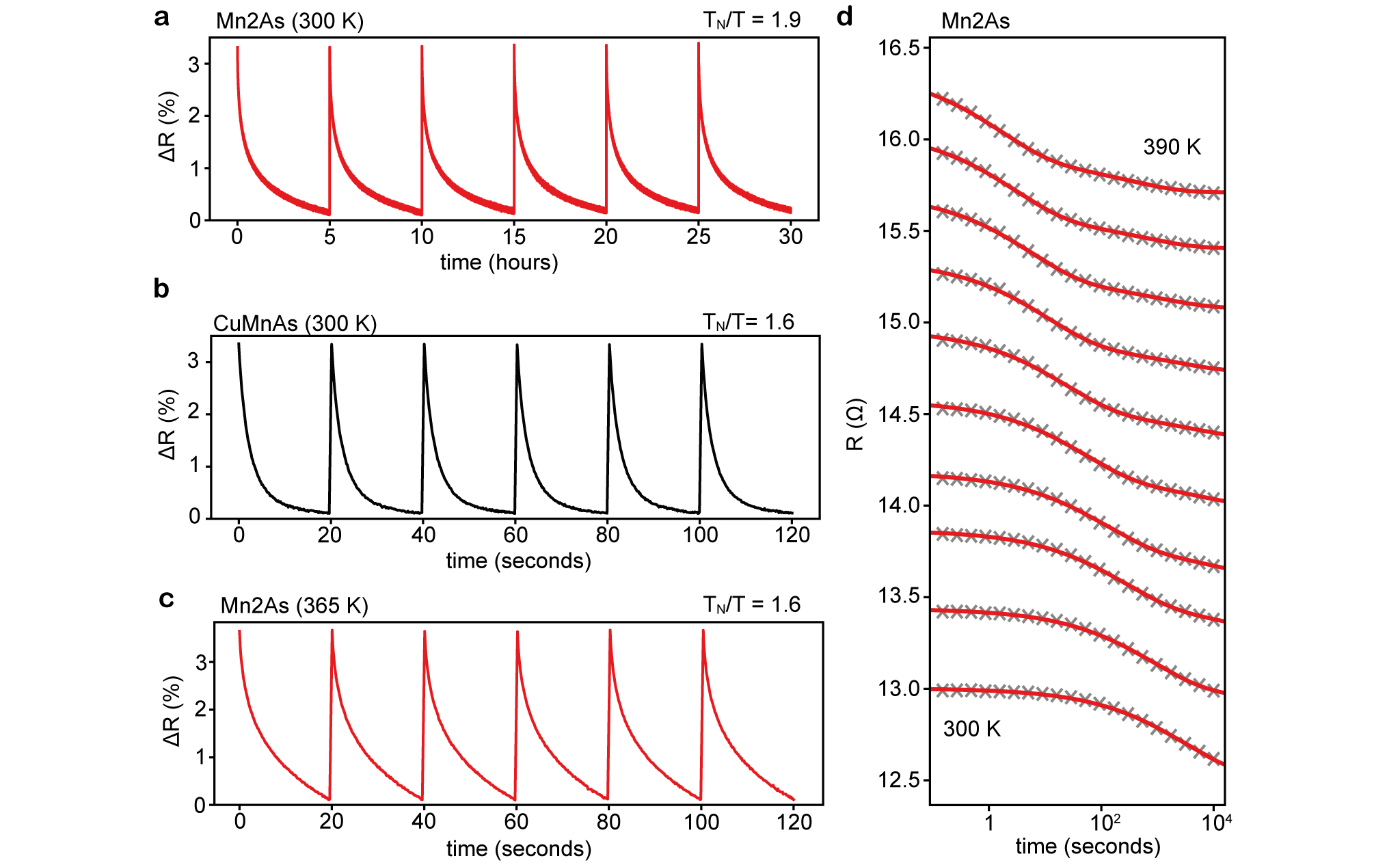}
\vspace{-0.4cm}
\caption{(a) Time-dependent resistance change induced by the quench-switching of Mn$_{2}$As by a sequence of six pulses, showing a characteristic resistance relaxation lasting for hours at 300K (corresponding to T$_{N}$/T = 1.9). (b) The same for CuMnAs, showing a characterising relaxation lasting for tens of seconds at 300K (corresponding to T$_{N}$/T = 1.6). c) The same for Mn$_{2}$As, showing a characteristic relaxation lasting for tens of seconds at 365K (corresponding to T$_{N}$/T = 1.6). (d) Time-dependent resistance change induced by the quench-switching of Mn$_{2}$As at temperatures ranging from 300 to 390 K in steps of 10 K, highlighting the temperature dependency of the relaxation process. The red lines correspond to fit to the data by stretch-exponential functions.}
\label{fig3}
\end{figure*}


From XRD radial measurements we extract c = 6.25~\AA $\,$ and a = 3.80~\AA. Rotated by 45 degree with respect to the substrate, Mn$_2$As is coincidence-site lattice matched to GaAs, which means that the [100] direction of Mn$_2$As aligns with the [110] direction of GaAs, in analogy to CuMnAs deposited on III-V substrates. Remaining mismatch is approximately 5\%.


To assess the crystallographic structure and quality, in Fig.~\ref{fig1}c we show the XRD radial scans of the Mn$_2$As film grown on GaAs substrate, together with data for the reference CuMnAs grown on GaAs and GaP. For both Mn$_2$As and CuMnAs, XRD confirms the formation of pure tetragonal phase with similar lattice constants. Our Mn$_2$As epilayers are of high quality, which is apparent from clear thickness fringes around all the 001, 002, 003 and 004 peaks. The XRD reciprocal maps and STEM measurements further reveal that the Mn$_{2}$As films have superior quality compared to the CuMnAs films in terms of mosaicity and that they are fully relaxed. The relaxation occurs via formation of pairs of misfit dislocations of opposite Burgers vectors at the interface with the GaAs substrate \cite{narayan2002formation}. The overall high quality of the Mn$_2$As films, corroborated by scanning transmission electron microscopy (STEM) measurements in Fig. \ref{fig1}d and the Supplemental material,  is mainly due to the absence of point and extended crystallographic defects, which commonly form in CuMnAs \cite{krizek2020molecular}. Remarkably, even though Mn$_{2}$As bulk lattice constant is closer to GaP than to GaAs, we found that growth on GaP results in formation of extended defects and chemical intermixing at the interface. In the rest of the paper we, therefore, focus on Mn$_2$As films grown on GaAs (more detailed STEM comparison and reciprocal map analysis are given in the supplemental material).

In order to confirm the expected compensated antiferromagnetic ordering of Mn$_{2}$As thin films \cite{yuzuri1960magnetic,austin1962magnetic,yang1998electronic}, we performed SQUID measurements and compared the results to ferromagnetic MnAs thin film, as shown in Fig. \ref{fig2}a. At 10~K the Mn$_{2}$As films show only negligible magnetic moment, inspite of approximately 50\% higher density of Mn atoms compared to MnAs. 

Similar to CuMnAs \cite{zubavc2021hysteretic,wang2020spin}, the AFM order strongly influences magneto-transport properties of our Mn$_2$As films. Magnetoresistance (MR) of the order of 0.1\% is observed, see Fig.~\ref{fig2}b, which depends on the direction of applied in-plane magnetic field. Saturation of a spin-flop-like reorientation in field above $\approx 6$~T marks the upper limit of a range where the MR is dominated by magnetic moments changing their direction. For stronger fields, the longitudinal and transversal MR (i.e. blue and green curves) run nearly parallel to each other and they are dominantly of orbital origin \cite{Zhang2019}. The difference between the two values of $R_{xx}$ (approximately 0.5\%) is then the anisotropic MR (AMR) which is of similar magnitude \cite{Ritzinger2023} as in transition metals. The sign of the AMR in Mn$_2$As films is opposite to that in CuMnAs thin films \cite{zubavc2021hysteretic}, and the same as in CuMnAs bulk \cite{Volny2020}). Higher spin-flop field in Mn$_2$As is consistent with the higher Néel temperature.



\section{Quench switching}

To perform the quench-switching experiments selected Mn$_{2}$As samples were patterned into bars with one-square (40$\times$40~$\mu$m) channel constriction in the middle. Electrical pulses of 10 to 20~V amplitude and 100~$\mu$s duration were applied to induce the quench-switched state. The four-probe resistance was recorded after each pulse, exhibiting a characteristic and reproducible relaxation shown in Fig.~\ref{fig3}a as the relative change $\Delta R$. We note, that the heat generated by the switching pulses dissipates into the substrate and the sample holder significantly faster than the shortest time resolution used within this study\cite{surynek2024picosecond}.

In Fig.~\ref{fig3}a and b we compare the quench-switching characteristics of Mn$_{2}$As and CuMnAs at room temperature. For each material we adjusted the pulse amplitude so that it lead to $\Delta R \sim$ 3 $\%$. The device response to pulsing is in both cases qualitatively similar. An important difference is that at the base temperature of 300 K we observe relaxation on the time scales of hours in the case of Mn$_2$As and seconds in the case of CuMnAs.

The relaxation time of the quench-switched resistance in Mn$_{2}$As strongly depends on the base temperature, similarly to CuMnAs \cite{kavspar2021quenching}. In Fig.~\ref{fig3}c we show results of a measurement of the same Mn$_2$As device as in Fig.~\ref{fig3}a, but at the base temperature of 365 K. The magnitude of the quench-switched signal is set to be the same as at 300 K by adjusting (decreasing) the pulse amplitude. The character of the relaxation remains the same as at 300~K, but the relaxation time shifts to the order of tens of seconds. This matches the relaxation time in CuMnAs at 300 K.

\begin{figure}[hbt!]
\vspace{0.2cm}
\includegraphics[scale=0.115]{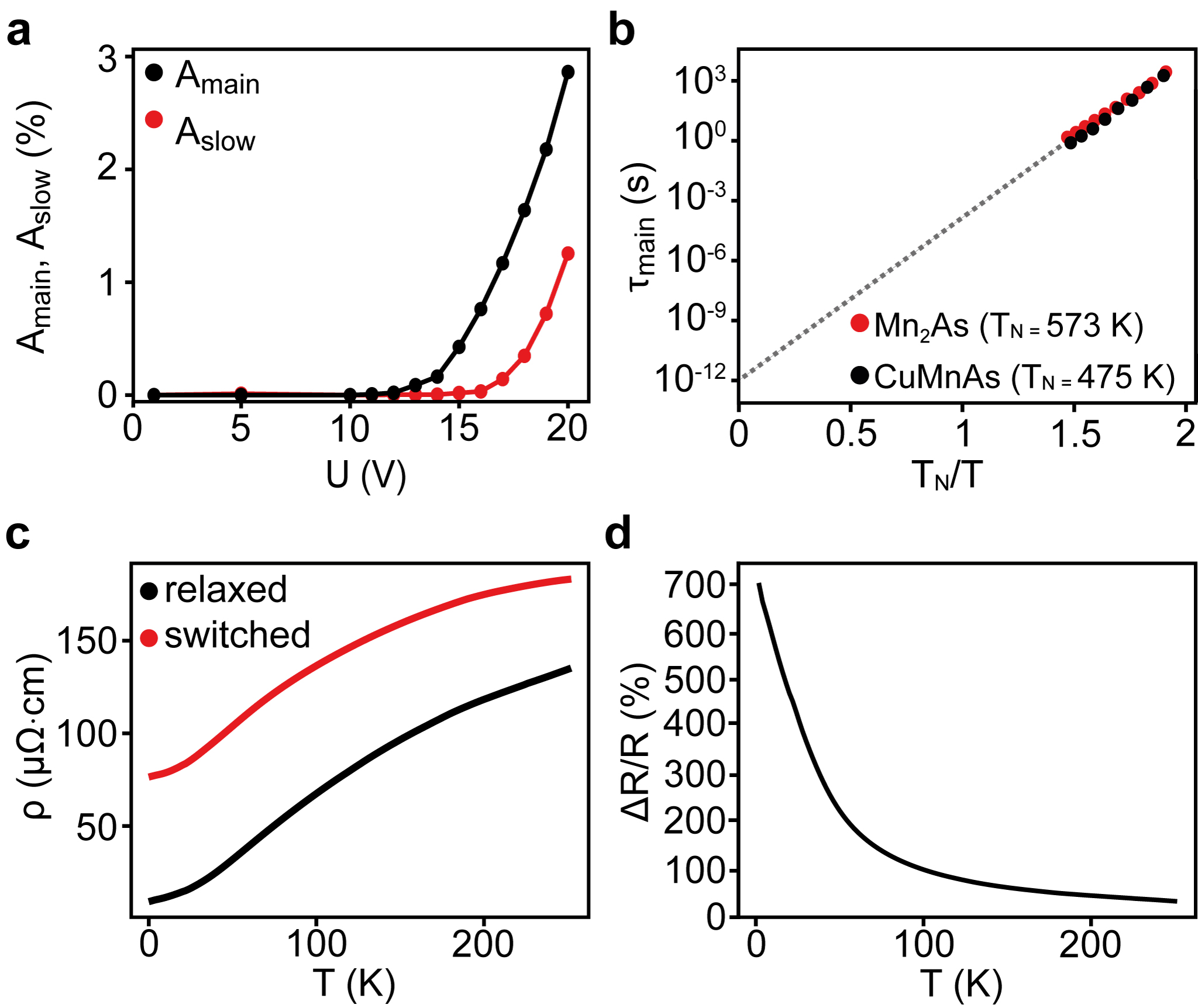}
\vspace{-0.4cm}
\caption{(a) Fitted amplitudes of the main and the slow components of the resistance relaxation of Mn$_2$As recorded at the base temperature of 370~K.
(b) Relaxation times of the main component in Mn$_2$As and CuMnAs plotted as a function of inverse temperature normalized to T$_N$.
(c) Temperature dependence of resistance of a Mn$_{2}$As device in the maximally quench-switched state and in the relaxed state; above $\sim$250~K the quench-switched state starts to relax (not shown in the figure).
(d) The same data as in (c) plotted as the relative change in the resistance (with respect to the relaxed value).}
\label{fig4}
\end{figure}

In Fig.~\ref{fig3}d we plot the recorded relaxations of the quench-switched signal in Mn$_{2}$As on the logarithmic time-scale for different base temperatures ranging from 300~K to 390~K. Similarly to CuMnAs, in the measured time-window of $10^{-1}$ -- $10^4$~s the data can be fitted with two Kohlraush stretched exponential functions \cite{kavspar2021quenching},
\begin{equation*}
\Delta R(t) = A_{main} \cdot e^{-(\frac{t}{\tau_{main}})^\beta} + A_{slow} \cdot e^{-\left(\frac{t}{\tau_{slow}}\right)^\beta} 
\end{equation*}
where ${\beta}$ is the stretching parameter. The best fits were obtained with ${\beta}$=0.5. The two exponentials reflect the presence of two relaxation components with different characteristic relaxation times $\tau_{main}$ and $\tau_{slow}$ and amplitudes $A_{main}$ and $A_{slow}$. 
Although the physical origin of these components remains unclear, their amplitudes depend in a threshold-like manner on the amplitude of the heating pulse. This is shown in Fig.~\ref{fig4}a, where an earlier onset of the main component and its prevalence over the slow component can be seen.
To allow for comparison between Mn$_{2}$As and CuMnAs, in the further text we focus our analysis only on the main component which dominates the relaxation process in both cases. 

In Figure \ref{fig4}b we show the relaxation times of the main component for both Mn$_{2}$As and CuMnAs plotted against the inverse temperature. Their temperature dependencies can be fitted by an Arrhenius-like relation $\tau = \tau_0\cdot\exp(E_b/k_B T)$ \cite{meinert2018electrical}, where $\tau_0$ reflects the characteristic attempt rate and $E_b$ is the energy barrier governing the thermally activated relaxation. 
It can be seen that in both cases $\tau$ dependencies extrapolate towards the picosecond range for $1/T$ approaching 0, yielding picosecond $\tau_0$ expected for antiferromagnets.
Moreover, the dependencies for Mn$_2$As and CuMnAs clearly overlap, when the inverse temperatures in the Arrhenius plot are normalized by the respective Néel temperatures. The overlap implies that the energy barriers $E_b$ of the two materials also scale with the Néel temperature. 

Both materials can be also compared in terms of the maximum achievable magnitude of the quench-switching signal. In order to avoid current leakage of the high-amplitude ($\approx 30$~V) heating pulse into the GaAs substrate, the 20~nm thick Mn$_2$As film was grown on top of a AlGaAs buffer layer in this particular case.
Analogously to CuMnAs~\cite{kavspar2021quenching}, we measured the temperature dependency of the resistance in the quench-switched state and in the fully relaxed state. The Mn$_2$As device was quench-switched at 250~K to $\Delta R/R \approx 35\%$, which for the particular sample was close to the maximum at which still no structural damage was induced to the material. At 250~K (i.e.~at $T_N/T\approx 2.3$) the relaxation is practically inhibited and temperature dependence of the device resistance can be recorded down to 5~K while remaining in the fully quench-switched state (red curve in Fig.~\ref{fig4}c). Subsequently, the device was heated to 390~K, which triggered a relatively fast relaxation. After 4~hours at this temperature, the fully relaxed resistance was again recorded during the cool down back to 5~K (black curve in Fig.~\ref{fig4}c). The corresponding relative change in resistance is shown in Fig.~\ref{fig4}d and approaches 700$\%$ in Mn$_{2}$As at 5~K, compared to 100$\%$  previously reported in CuMnAs. Part of this difference comes from different residual resistivity ratios (RRR) of CuMnAs and Mn$_2$As. The highest RRR achieved in CuMnAs reaches about 6, in Mn2As we observed RRR as high as 24. We attribute the difference in RRR of the two materials to the absence of Mn/Cu antisite defects in Mn$_{2}$As. In the case of the device shown in Fig.~\ref{fig4}c) RRR reaches approx. 16. 


\section{Discussion}

Previous microscopic investigations of quench switching in CuMnAs \cite{kavspar2021quenching} revealed, that the resistance increase is associated with fragmentation of the magnetic state into domains with size less than $\approx 10$~nm. Later works connected the fragmentation to the presence of atomically sharp domain walls \cite{krizek2022atomically,zubavc2021hysteretic}, which would allow for the existence of magnetic domains with nm size that may remain unresolved by standard magnetic imaging methods. Formation and long-term stability of such domain walls seem to fall out of the framework of effective micromagnetic modeling, which predicts relatively wide (tens of nanometers) domain walls for the considered antiferromagnets. In the micromagnetic framework the domain wall width is given by an energy minimum associated with balance of competing exchange and anisotropy energies. In contrast to the micromagnetic approach, atomistic ab-initio simulations in CuMnAs revealed that a local minimum in the domain wall energy exists for the case of a 180$^\circ$ atom-to-atom spin rotation, i.e.~for the atomically sharp domain wall \cite{krizek2022atomically}. Energy barrier for the formation of such atomically sharp domain wall is directly related to the atom-to-atom exchange interaction.


In this work, we show by comparing the relaxation times in Mn$_2$As and CuMnAs that the energy barrier governing the quench-switching relaxation is proportional to the Néel temperature (see Fig.~\ref{fig4}b). Importantly, the scaling to the Néel temperature confirms the magnetic origin of the effect. Furthermore, since the Neel temperature is directly proportional to the exchange interaction, the scaling also represents additional support for the microscopic mechanism of quench switching based on atomically sharp domain walls whose energy is related solely to the strength of exchange interaction in contrast to the traditional micromagnetic domain mechanism sensitive also to the strength of magnetic anisotropy.



The stretching parameter of the relaxation relates to the dimensionality of the problem \cite{phillips2006axiomatic}. Interestingly, ${\beta}$=0.5 was found in Mn$_2$As while for CuMnAs the value 0.6 was observed, indicating 2D and 3D diffusion, respectively. This may be related to different spin stiffness along the c-axis in the two materials which affects the domain fragmentation in the c-direction. The difference of magnetic interactions along the c-direction is, in fact, of qualitative character: in CuMnAs the strongly antiferromagnetically coupled planes of Mn atoms are separated by non-magnetic planes of Cu atoms, whereas in Mn$_2$As the Cu planes are occupied by magnetically oriented Mn atoms and the magnetic unit cell is doubled. The difference in ${\beta}$ can also be related to differences in characteristic defects.

Mn$_{2}$As is of particular interest due to its high $T_N$. The associated temperature dependence of the quench-switching signal opens the possibility of performing a wider range of experiments aiming to characterize the quench-switched state such as the use of complex magnetic imaging techniques and even experiments studying relaxation dynamics at room temperature, where the slow relaxation process takes tens of hours. This is in contrast to CuMnAs, where lower $T_{N}$ results in impractically short relaxation times at room temperature for most microscopy techniques.

The relaxation time scale of quench switching in both materials is determined by $T/T_N$. Therefore, setting the correct base temperature range is essential for testing the presence of quench switching in other antiferromagnetic materials and for its separation from other effects. Changing the $T/T_{N}$ ratio also allows for precise tuning of the relaxation properties of quench-switching-based spintronic devices and tailoring their neuron-like long and short-term memory properties \cite{surynek2024picosecond}. 

\section{Conclusion}
We have shown that the recently reported intriguing quench-switching phenomenology is not limited solely to CuMnAs, but is a universal mechanism present in more antiferromagnetic materials, as demonstrated in this work in epitaxial Mn$_2$As. The comparison of quench switching in Mn$_{2}$As and CuMnAs shows clear scaling of the relaxation dynamics with the Néel temperature and confirms the magnetic origin of the effect. The experimental observation of the quench switching of resistivity in another material provides the evidence needed to establish and validate the theoretical description of the microscopic origin of the effect and its further refinement. 


\section{Acknowledgement} 
We thank M. Sousa (IBM) and IBM Research Zurich for the access to the TEM facility. The authors also acknowledge ScopeM for their support and assistance. The authors also acknowledge the support of GACR grant 21-28876J, MEYS grants LM2023051 and CZ.02.01.01/00/22$\_$008/0004594 and the Dioscuri Program LV23025 funded by MPG and MEYS. D.K. acknowledges the support from the Czech Academy of Sciences (project No. LQ100102201).

\bibliography{ms}

\makeatletter 
\renewcommand{\thefigure}{S\@arabic\c@figure}
\makeatother
\setcounter{figure}{0}

\section{Supplementary Information}

\subsection{Methods}

\subsubsection{XRD analysis}
The structure of the samples was analysed by high-resolution X-ray diffractometer equipped with high-flux 9 kW Cu rotating anode providing a monochromatic CuK$_{\alpha_{1}}$ beam, while using the D/teX Ultra 250 detector. The data were processed using xrayutilities Python package \citen{kriegner2013xrayutilities}.

\subsubsection{STEM sample preparation and measurement}
The lamellae of cross-sectional samples were prepared by Focused Ion Beam (FIB) (Helios 5 UX from Thermo Scientific) using AutoTEM 5 software (Thermo Scientific, the Netherlands) at ScopeM, ETH Zurich. A protective carbon layer was deposited on the selected region of interest first by an electron beam (2 kV, 13 nA) and subsequently by an ion beam (30 kV, 1.2 nA). The chuck milling and lamellae thinning were done at 30 kV with FIB current from 9 nA to 90 pA. Finally the lamellae were polished at 5 kV (21 pA) and finished at 2 kV (17 pA). The expected thickness was bellow 50 nm. 

The STEM images were acquired with a double spherical aberration-corrected JEOL ARM200F cold field emission gun scanning transmission electron microscope located at the Binnig and Rohrer Nanotechnology Center Noise-free laboratories at IBM Research Europe. The images were acquired at 200 kV.

The acquired STEM images were processed using FIJI - Fiji is just ImageJ freeware \cite{schindelin2012fiji}. Typically, the FFT pattern of the images was masked by a circular aperture and background noise was removed by subtracting a constant background.
The crystal simulations used for the STEM image overlays were made using the JP-Minerals VESTA visualization program for structural models \cite{momma2008vesta}.

\subsubsection{Device fabrication}
The hallbars with length of 20 $\mu$m and width of 40 $\mu$m were fabricated by laser lithography, using photoresis ma-P 1200. The protective AlOx capping layer is removed after exposition during development in ma-D 331. The mesa in Mn2As was wet etched in a in a solution of H3PO4.

\subsubsection{Quench-switching setup}
Voltage pulses of desired lengths were generated by a laboratory DC source connected to a fast-switching transistor.
National Instruments DAQ card 6356 was used to open and close the transistor gate, as well as to measure two-probe and four-probe resistance of the sample.
Magnetoresistance measurements up to 14 T were performed in commercial PPMS system from QD.
Temperature controlled measurements were performed in commercial Oxford Instruments helium cryostat.

\begin{figure*}[hbt!]
\vspace{0.2cm}
\includegraphics[scale=0.2]{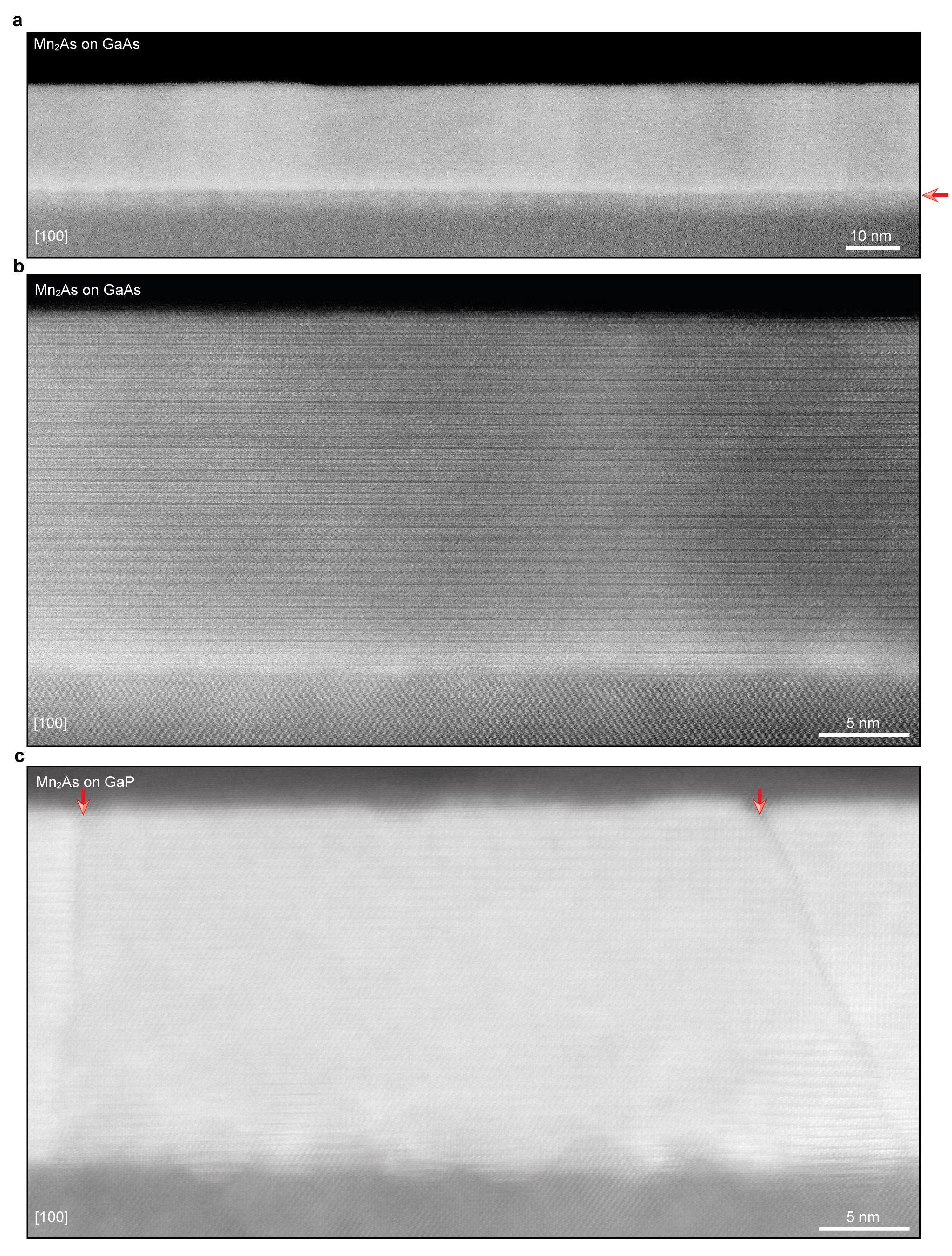}
\vspace{-0.4cm}
\caption{(a) HAADF-STEM image of 20 nm thick Mn$_{2}$As on GaAs showing overall homogenity of the layer and hints of periodic contrast corresponding to misfit dislocations at the interface (red arrow). (b) HAADF-STEM zoom in on the same film. (c) HAADF-STEM image of 20 nm thick Mn$_{2}$As on GaP showing presence of extended defects (red arrow) and hints of chemical interaction at the interface.}
\label{figSTEM}
\end{figure*}

\begin{figure*}[hbt!]
\vspace{0.2cm}
\includegraphics[scale=0.2]{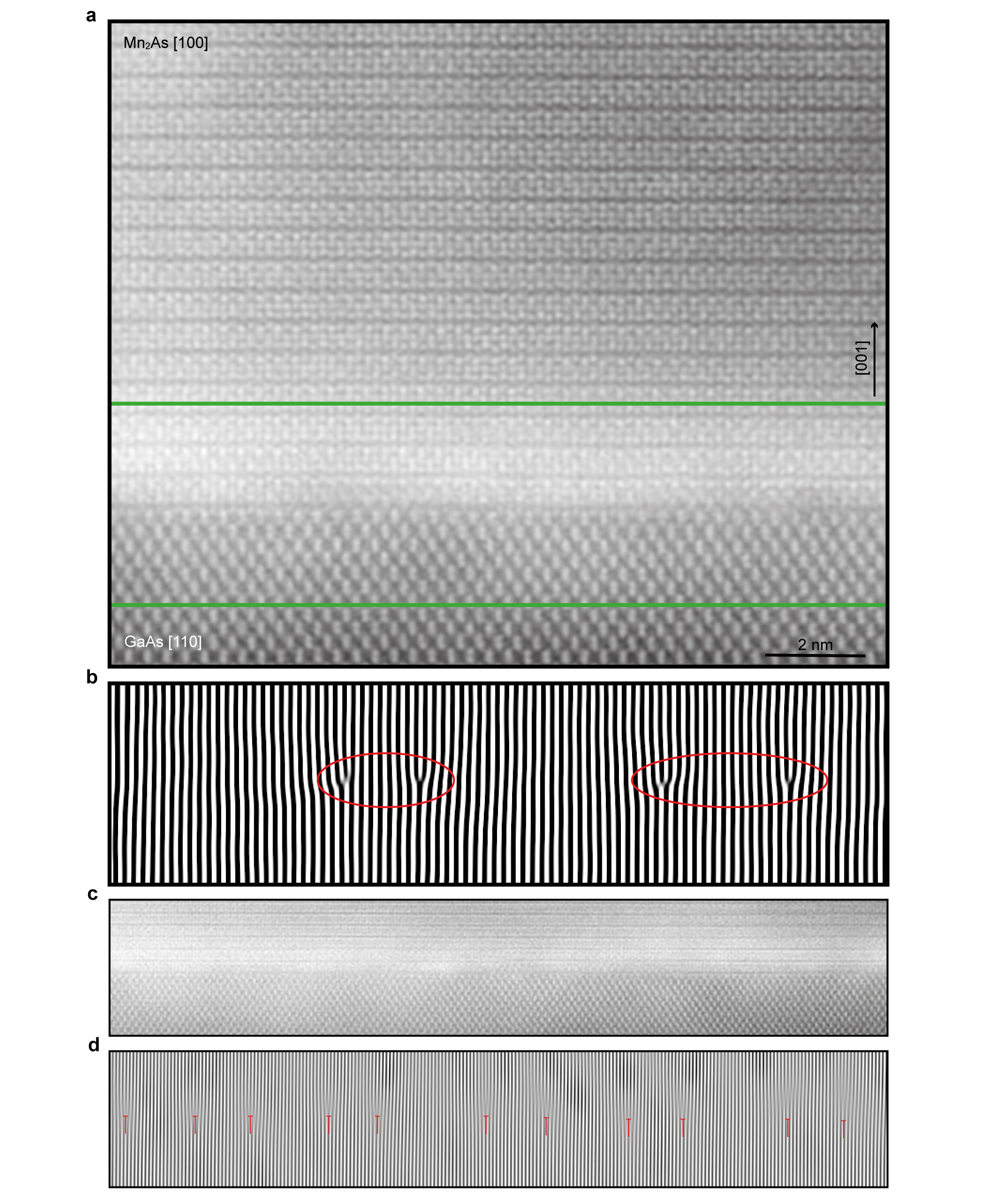}
\vspace{-0.4cm}
\caption{(a) HAADF-STEM image of a portion of 20 nm thick Mn$_{2}$As on GaAs showing details of the interface between the film and the substrate. (b) Bragg filtered HAADF-STEM of the area highlighted in (a) by green, showing a-periodic pairs of misfit dislocation at the interface. (c) and (d) are the same as a) and b), but for larger portion of the interface.}
\label{figSTEM2}
\end{figure*}

\begin{figure*}[hbt!]
\vspace{0.2cm}
\includegraphics[scale=0.25]{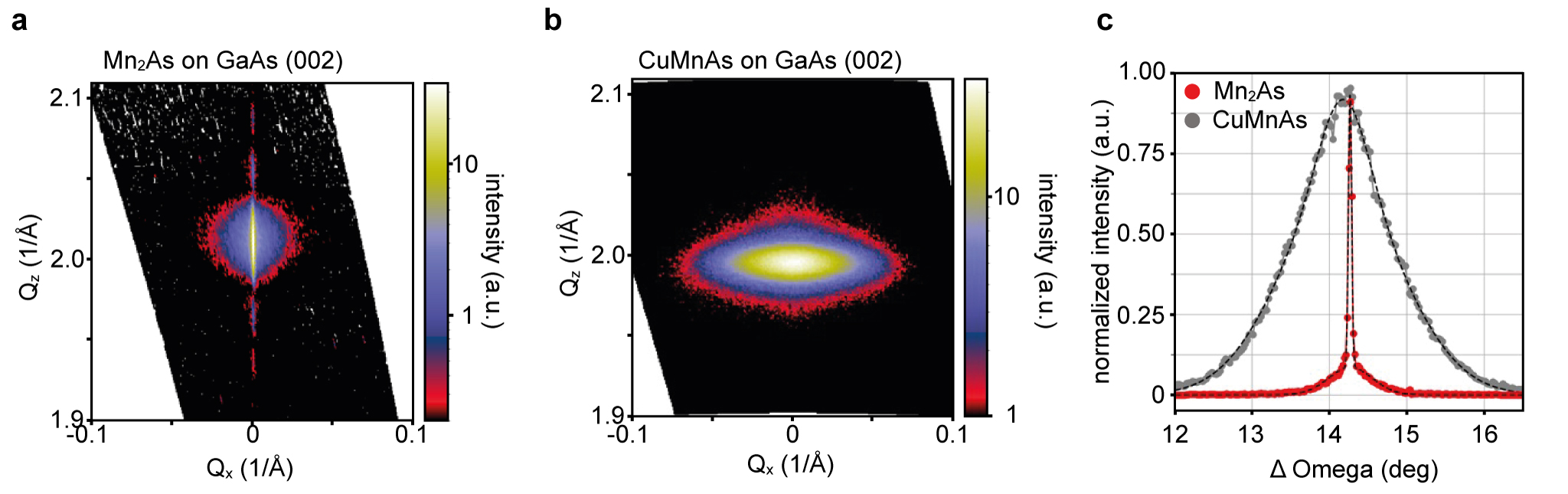}
\vspace{-0.4cm}
\caption{(a) Reciprocal maps of 002 peaks of 20 nm thick Mn$_{2}$As on GaAs and CuMnAs on GaAs. (c) Linecuts through the maps in a) and b) showing that higher quality of Mn$_{2}$As, in terms of very narrow peak and subtle diffusive background, likely corresponding to missfit dislocation arrays.}
\label{figSXRD}
\end{figure*}

\begin{figure*}[hbt!]
\vspace{0.2cm}
\includegraphics[scale=0.25]{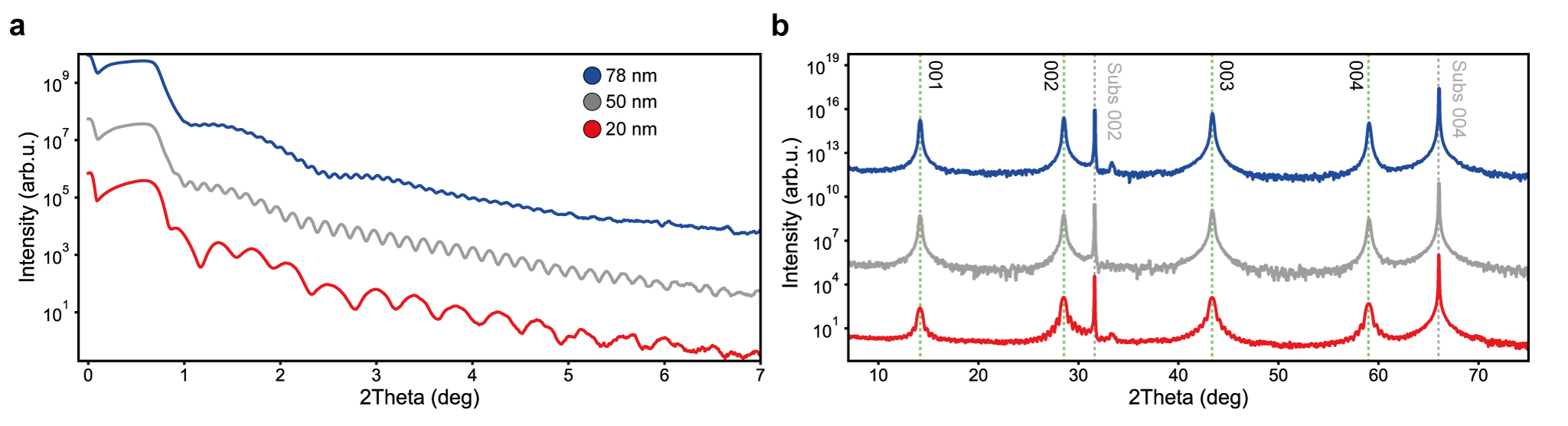}
\vspace{-0.4cm}
\caption{(a) X-ray reflectivity measured on 20, 50 and 78 nm thick Mn$_{2}$As film grown on GaAs substrate, demonstrating smooth surface of the films at different thicknesses. (b) XRD radial scans of the same smaples, demonstrating high quality and absence of emergence of a parasitic phase (such as MnAs etc.) at different thicknesses. We note that the relatively low intensity side peak at 33 degrees, is likely related to the large spotsize of the XRD machine. As shown in Fig. \ref{SRHEEDSEM} the ideal stoichiometry region is limited due to As flux distribution in our MBE machine.}
\label{figSthickXRD}
\end{figure*}

\begin{figure*}[hbt!]
\vspace{0.2cm}
\includegraphics[scale=0.25]{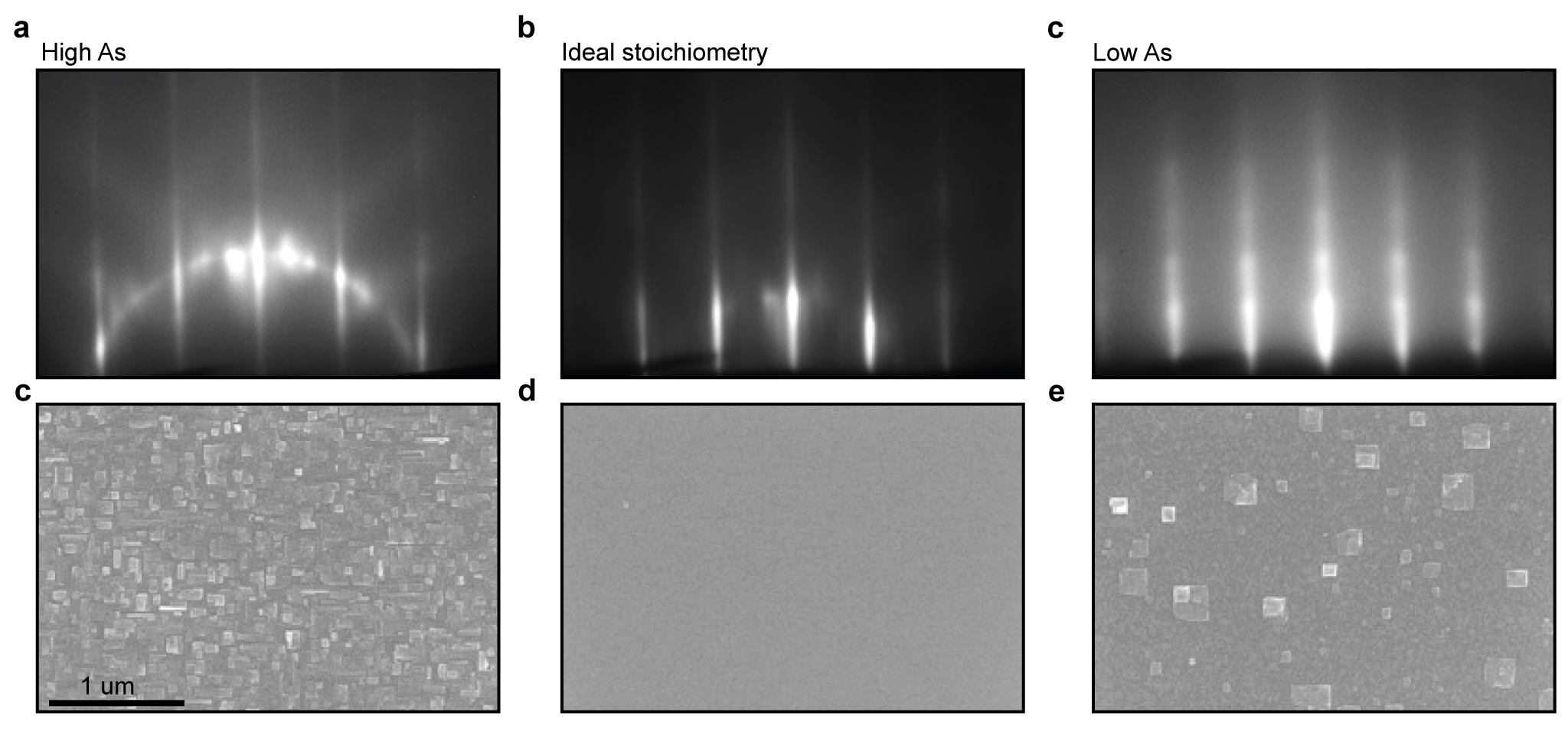}
\vspace{-0.4cm}
\caption{(a), (b) and (c) RHEED snapshots along the [100] direction of Mn$_{2}$As right after growth for high As flux, ideal stoichiometry and low As regions, respectively. (e), (d) and (f) corresponding SEM images showing the typical surface features. The geometry of our MBE system results in a significant As flux distribution over a 2inch wafer, even when the substrate is rotated during growth. This typically results in a ring of width between 3-6 mm of the ideal surface shown in b), around the 2inch wafer. The advantage of the As distribution is, on the other hand, that the growth is partially self-calibrated with respect to As and also small daily changes in Mn flux. If fluxes change, the ring with ideal surface simply shifts over the 2inch and its location is typically easily visible by eye or in SEM after growth.}
\label{SRHEEDSEM}
\end{figure*}

\end{document}